\documentclass[preprint,prl,aps,amsmath,amssymb,color,superscriptaddress]{revtex4}
\usepackage{stmaryrd}
\usepackage{amsmath}
\usepackage{amssymb}
\usepackage{graphicx}
\usepackage{dcolumn}
\usepackage{bm}
\usepackage{tabularx}
\usepackage{color}
\begin{document}
\begin{titlepage}
\title{One-Dimensional Magnetic Excitonic Insulators}
\author{Jing Liu}
\affiliation{Key Lab of advanced optoelectronic quantum architecture and measurement (MOE), and Advanced Research Institute of Multidisciplinary Science, Beijing Institute of Technology, Beijing 100081, China}
\affiliation{These authors contribute equally to this work.}
\author{Hongwei Qu}
\affiliation{Key Lab of advanced optoelectronic quantum architecture and measurement (MOE), and Advanced Research Institute of Multidisciplinary Science, Beijing Institute of Technology, Beijing 100081, China}
\affiliation{These authors contribute equally to this work.}
\author{Yuanchang Li}
\email{yuancli@bit.edu.cn}
\affiliation{Key Lab of advanced optoelectronic quantum architecture and measurement (MOE), and Advanced Research Institute of Multidisciplinary Science, Beijing Institute of Technology, Beijing 100081, China}
\date{\today}

\begin{abstract}
Dimensionality significantly affects exciton production and condensation. Despite the report of excitonic instability in one-dimensional materials, it remains unclear whether these spontaneously produced excitons can form Bose-Einstein condensates. In this work, we first prove statistically that one-dimensional condensation exists when the spontaneously generated excitons are thought of as an ideal neutral Bose gas, which is quite different from the inability of free bosons to condense. We then derive a general expression for the critical temperature in different dimensions and find that the critical temperature increases with decreasing dimension. We finally predict by first-principles $GW$-BSE calculations that experimentally accessible single-chain staircase Scandocene and Chromocene wires are an antiferromagnetic spin-triplet excitonic insulator and a ferromagnetic half-excitonic insulator, respectively.
\end{abstract}

\maketitle
\draft
\vspace{2mm}
\end{titlepage}

In the 1960s, theoretical physicists propose a new state of matter called the excitonic insulator, which has a many-body ground state similar to the superconducting state\cite{Mott,Knox,Kohn,Rice}. Its onset is accompanied by spontaneous symmetry breaking, may leading to interesting physical phenomena such as superfluidity\cite{Eisenstein,Jiangtri} and electron ferroelectricity\cite{Sham,Batista}. Intertwining with magnetic and topological order also gives rise to novel quantum states, including single-spin exciton condensates\cite{Jianghf}, spin-triplet excitonic insulators\cite{Jiangtri,Liufeng,Mazzone} and topological excitonic insulators\cite{Durr,Daniele,Wusanfeng,Cobden,DongTEI,LiXZ}. However, to date, there is no universally accepted evidence that excitonic insulators exist in real-life materials.

The formation of excitonic insulators involves two aspects: (1) excitonic instability, i.e., spontaneous generation of excitons when the exciton binding energy exceeds the one-electron excitation gap, and (2) exciton condensation, i.e., the macroscopic occupation of the ground state by these spontaneously generated excitons after undergoing a quantum statistical phase transition. Both aspects are sensitive to dimensionality, but the dependence tends to be opposite. On the one hand, dimensionality reduction weakens the electron-hole screening effect and enhances their binding, thus promoting exciton generation. As such, low-dimensional excitonic insulators have the potential to outperform the traditional material choices of small-gap semiconductors or small-band-overlap semimetals. First-principles studies have indeed shown that excitonic instability is still possible in two-dimensional (2D) semiconductors with gaps up to $\sim$3 eV\cite{Jianghf,Jiangtri}. It is for this reason that the rise of 2D materials in recent years has renewed interest in excitonic insulators\cite{usEI,usDong,DongAlSb,Daniele,Wusanfeng,Cobden,LxzNJP,DongTEI,LiXZ}. On the other hand, however, exciton condensation in low dimensions seems to be ruled out by the Hohenberg-Mermin-Wagner theorem\cite{Hohenberg,Mermin}, which states the absence of long-range order in 1D and 2D homogeneous systems at non-zero temperature due to quantum or thermal fluctuations. Based on a similar application of the Bogoliubov inequality, this was subsequently extended by Walker to the nonexistence of 1D and 2D excitonic insulators\cite{Walker}. It is worth noting that all these proofs were obtained under idealized conditions, e.g., in the thermodynamic limit. With the continuous progress of low-dimensional physics, there have been different levels of theoretical models and experimental approaches that go beyond the limits of the Hohenberg-Mermin-Wagner theorem, as exemplified by the realization of stable ferromagnetic order in 2D Cr$_2$Ge$_2$Te$_6$\cite{Gong} and CrI$_3$\cite{Huang}.

In this work, we focus on 1D excitonic insulators. Physically, 1D quantum many-body systems are of great scientific importance because theoretical descriptions of systems of interacting particles are very difficult and, in some cases, exact solutions exist only in one dimension. Yet, 1D excitonic insulators have historically received little attention\cite{RiceJPCM,Rice9,Rice11,VarsanoNC,Hellgren,MnCp,CrBz,Barborini} compared to 2D ones. In terms of material platforms, excitonic instabilities have been reported for conjugated polymer\cite{Rice9,Rice11}, carbon nanotubes\cite{VarsanoNC,Hellgren} and organometallic wires\cite{MnCp,CrBz}. In terms of Bose-Einstein condensation, by mapping to the Heisenberg model\cite{Batista}, a previous study\cite{Batista4} showed that the 1D excitonic phase is critical (i.e., there is no long-range order). Hence, there is a case that 1D exciton condensation is ruled out by the Hohenberg-Mermin-Wagner theorem. But in practice it is generally accepted that the finite size and dimension of real-life materials will weaken the statement of the Hornberger-Mermin-Wagner theorem, and thus the existence of long-range order (e.g., superconductivity, magnetism, and Bose-Einstein condensation) in confined geometries remains under debate. To the best of our knowledge, a more fundamental question of whether 1D condensation can occur when considering spontaneously generated excitons as an ideal neutral Bose gas has not yet been addressed. It is well known that ideal free bosons are not allowed to undergo 1D condensation, which makes the study of non-interacting case even more scientifically interesting.

Consider a free boson gas of mass $m$ and momentum $p$ for a 1D system of length $L$. The number of quantum states in the energy range $\varepsilon$ to $\varepsilon$+d$\varepsilon$ is
\begin{equation} \label{eq1}
D\left( \varepsilon  \right){\rm d}\varepsilon  = \frac{{2L}}{h}{\left( {\frac{m}{{2\varepsilon }}} \right)^{\frac{1}{2}}}{\rm d}\varepsilon.
\end{equation} Combining Eq. (1) with the Bose-Einstein distribution functions, the total number $N$ of bosons is given by
\begin{equation}
\label{eq2}
    N = \int_{\rm{0}}^\infty  {\frac{{D(\varepsilon ){\rm d}\varepsilon }}{{{e^{\beta (\varepsilon  - \mu )}} - 1}}} = \frac{L}{\pi }\int_0^\infty  {\frac{{{\rm d}k}}{{{e^{\beta (\varepsilon  - \mu )}} - 1}}},
\end{equation}
where ${\beta}={\left( {{k_{\rm B}}{T}} \right)^{ - 1}}$. This integral diverges if chemical potential $\mu$ = 0. Physically, the divergence means that the continuum states can accommodate infinite particles. Therefore the condensate is excluded for a 1D free boson gas since its $\mu$ is zero.

However, the situation is different for exciton condensation, either non-equilibrium or spontaneous. According to statistical mechanics, $\mu$ equals the change in Gibbs free energy at constant temperature and pressure. The production of excitons in solids, no matter by photoexcitation or spontaneous formation, can be approximated as an isothermal and isobaric process. Therefore, $\mu$ is roughly equivalent to the exciton formation energy. For non-equilibrium condensation, exciton production is energy-consuming, hence $\mu >$ 0. For excitonic insulators, spontaneous production of excitons implies that the formation energy is negative, hence $\mu <$ 0. As such, the divergent behavior of Eq. (2) will be eliminated for exciton condensation.

Because it is the excitons that constitute the ground state of excitonic insulators, there are discrete exciton bands below the continuous single-particle states. Without losing generality, let us assume that the lowest exciton energy is $\varepsilon _0 < 0$. At this point, the upper limit of the $\mu$ in Eq. (2) is $-\lvert \varepsilon_0\rvert$.

Now the maximum particle number $N_{\rm c}$ that can be accommodated in the continuum spectrum is
\begin{equation}
\label{eq3}
	{N_{\rm c}} = \frac{L}{\pi }\int_0^\infty  {\frac{{\rm d}k}{{{e^{\beta \left( {\varepsilon  + \left| {{\varepsilon _0}} \right|} \right)}} - 1}}} \\
	= A\sum\limits_{j = 1}^\infty  {\frac{{{{({e^{ - \beta |{\varepsilon _0}|}})}^j}}}{{{j^{\frac{1}{2}}}}}} \\
    < A\sum\limits_{j = 1}^\infty  {({e^{ - \beta |{\varepsilon _0}|}})}^j \\
    = A\frac{1}{e^{ \beta |{\varepsilon _0}|}-1},
\end{equation}
where $A=\sqrt {{mL^2}/{2 \pi {\hbar^2} \beta}}$. Definitely, the $N_{\rm c}$ has an upper limit. When the number of preformed excitons $N$ exceeds this critical value, the excess particles ($N-N_{\rm c}$) can only spill over into the discrete states. Imagine the extreme case where there is only one discrete state. Once $N > N_{\rm c}$, the excess excitons will inevitably condense into this state and form a macroscopic occupation of the ground state. Therefore, 1D excitonic insulators do exist when considering the spontaneously generated excitons as an ideal neutral Bose gas.

In this sense, one may evaluate the critical temperature $T_{\rm c}$ from the $N_{\rm c}$ as a function of the 1D exciton density $n_{\rm {1}}$, the exciton mass $m$, and the lowest exciton level $\varepsilon_0$,
\begin{equation}
\label{eq4}
	{n_{\rm {1}}} = \frac{N_{\rm c}}{L}\\
	= (\frac{mk_{\rm B}T_{\rm c}}{{2 \pi \hbar^2}})^{\frac{{\rm{1}}}{2}}\sum\limits_{j = 1}^\infty  {\frac{{{{({e^{ - |{\varepsilon _0}|/{k_{\rm B}}{T_{\rm c}}}})}^j}}}{{{j^{\frac{1}{2}}}}}}.
\end{equation}
Likewise, we derive the $T_{\rm c}$ for 2D and 3D excitonic insulators as in Eqs. (5) and (6)
\begin{equation}
\begin{split}
\label{eq5}
{n_{\rm {2}}} = \frac{{ mk_{\rm B}T_{\rm c}}}{{2\pi {\hbar ^2}}}\sum\limits_{j = 1}^\infty  {\frac{{{{({e^{-|{\varepsilon_0}|/k_{\rm B}T_{\rm c}}})}^j}}}{{{j}}}},
\end{split}
\end{equation}
\begin{equation} \label{6}
{n_{\rm {3}}} = (\frac{mk_{\rm B}T_{\rm c}}{{2 \pi \hbar^2}})^{\frac{{\rm{3}}}{2}}\sum\limits_{j = 1}^\infty  {\frac{{{{({e^{ -|{\varepsilon_0}|/k_{\rm B}T_{\rm c}}})}^j}}}{{{j^{\frac{{\rm{3}}}{2}}}}}}.
\end{equation}
Writing Eqs. (4), (5) and (6) in a unified form gives the $T_{\rm c}$ for $D$-dimensional excitonic insulators
\begin{equation} \label{7}
{n_{D}} = (\frac{mk_{\rm B}T_{\rm c}}{{2 \pi \hbar^2}})^{\frac{{D}}{2}}\sum\limits_{j = 1}^\infty  {\frac{{{{({e^{ -|{\varepsilon_0}|/k_{\rm B}T_{\rm c}}})}^j}}}{{{j^{\frac{{D}}{2}}}}}}.
\end{equation}

\begin{figure}[tbp]
	\includegraphics[width=1\columnwidth]{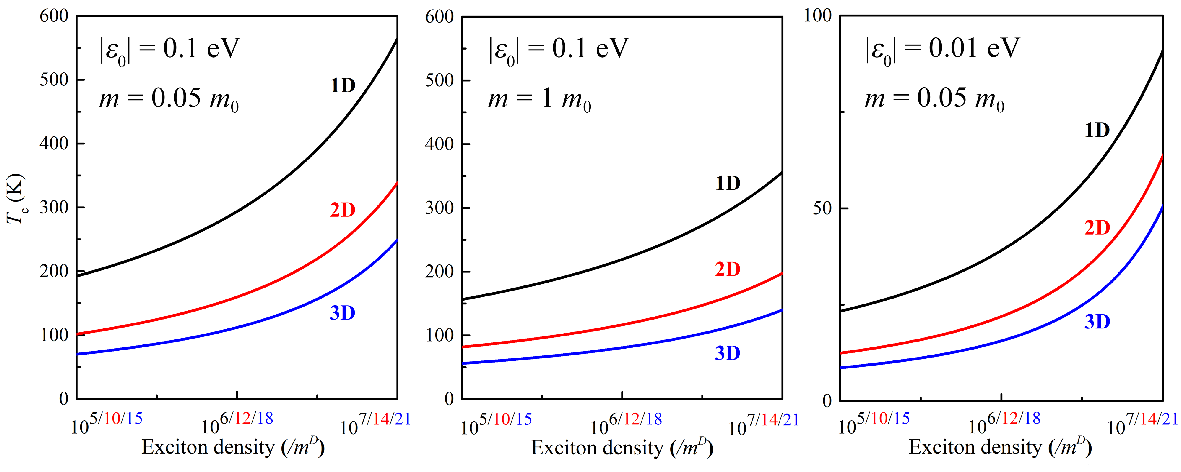}
	\caption{\label{Fig1} (Color online) The critical temperature $T_c$ in different dimensions as a function of normalized exciton density $n_{D}$, with the lowest exciton level $\lvert \varepsilon_0\rvert$ and the exciton mass $m$ set to $\lvert \varepsilon_0\rvert$ = 0.1 eV and $m$=0.05 $m_0$ (Left), $\lvert \varepsilon_0\rvert$ = 0.1 eV and  $m$=$m_0$ (Middle), and $\lvert \varepsilon_0\rvert$ = 0.01 eV and $m$ = 0.05 $m_0$ (Right). Here $m_0$ is the mass of free electrons. Different colours correspond to different dimensions.}
\end{figure}

In Fig. 1, we show the $T_{\rm c}$ of different dimensions as a function of the normalized $n_{D}$, as well as the effects of $\lvert \varepsilon_0\rvert$ and $m$. Three interesting trends are found. First, under the same conditions, the $T_{\rm c}$ presents the order 1D $>$ 2D $>$ 3D. Therefore, a decrease in dimensionality favours an increase in the $T_{\rm c}$. Second, $\lvert \varepsilon_0\rvert$ and $m$ have opposite effects on the $T_{\rm c}$. Increasing the former increases the $T_{\rm c}$, but increasing the latter decreases the $T_{\rm c}$. Third, the effect of $\lvert \varepsilon_0\rvert$ is more significant in comparison. For example, at $m$=0.05 $m_0$, decreasing $\lvert \varepsilon_0\rvert$ by a factor of ten leads to an order of magnitude decrease in the $T_{\rm c}$. In contrast, at $\lvert \varepsilon_0\rvert$=0.1 eV, a twenty-fold increase in $m$ causes a much smaller decrease in the $T_{\rm c}$.

Next we turn to consider material realization. Candidates for 1D excitonic insulators are very rare, and so far only conjugated polymer\cite{Rice9,Rice11}, carbon nanotubes\cite{VarsanoNC,Hellgren} and organometallic molecular wires\cite{MnCp,CrBz} have been theoretically predicted to exhibit excitonic instabilities. The recent synthesis of single-chain staircase ferrocene FeCp$_2$ nanowires (FeNWs) up to 50 nm in length not only breaks the current barrier for the length of such organometallic wires, but also pave the way for the realization of similar metallocene wires (TMNWs) with different embedded transition metals\cite{AFM}. In this work, we focus on Scandocene (ScNW) and Chromocene (CrNW) wires, which can be regarded as ScCp$_2$ and CrCp$_2$ molecules linked by the dehydrogenation of cyclopentadiene (Cp) rings (See Fig. 2).

\begin{figure}[tbp]
	\includegraphics[width=0.9\columnwidth]{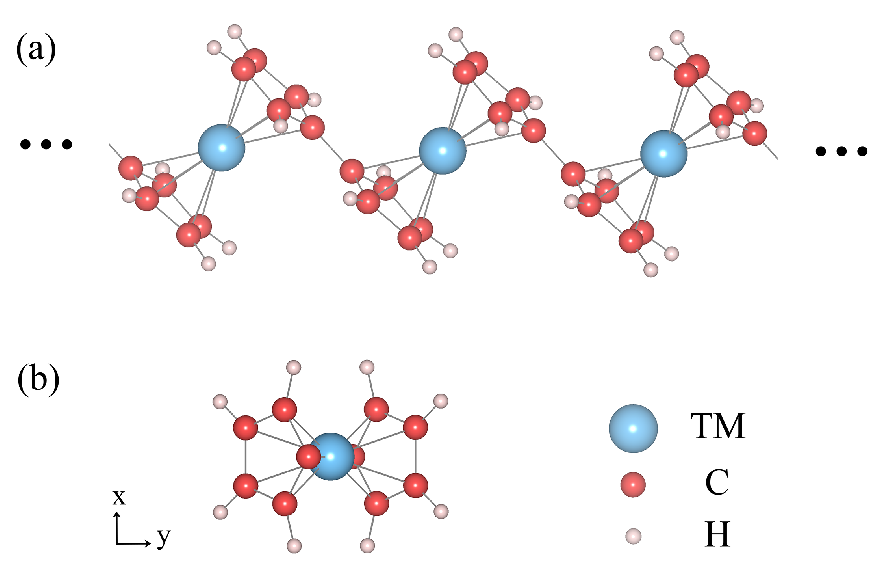}
	\caption{\label{Fig2} (Color online) (a) Side and (b) top views of a staircase metallocene nanowire as derived from the experiment\cite{AFM}. It is composed of staggered sandwich metallocene molecules (TMCp$_2$) linked by the dehydrogenation procedure, where one cyclopentadiene (Cp) ring is rotated by $36^\circ$ with respect to the other.}
\end{figure}

Density-functional-theory calculations were performed within the Perdew-Burke-Ernzerhof (PBE)\cite{PBE} functional as implemented in the Vienna ab initio Simulation Package (VASP)\cite{vasp}. The electron-ion interaction was described by a projector-augmented-wave method\cite{PAW,PAW2} with an energy cutoff of 450 eV. A $1 \times 1 \times 45$ \emph{k}-mesh was used to sample the Brillouin zone. A vacuum layer of at least 10 \AA\ was added to separate two neighboring wires. All atomic positions are fully relaxed without any symmetry constraint until the residual forces are less than 0.01 eV/\AA. The single-shot $G_0W_0$ method\cite{Hybertsen} was employed for the quasi-particle band structures, with a \emph{k}-grid of $1 \times 1 \times 19$ and 240 bands within our computational capabilities. The Bethe-Salpeter equation (BSE) was solved on top of the $GW$ results for excitonic properties\cite{Rohlfing}, including two valence and four conduction bands to build the Hamiltonian. Spin-orbit coupling was ignored because our test calculations show that it does not change the conclusion.

\begin{table*}
	\caption{Optimized lattice constant in \AA\ ($c$), relative formation energy with FeNW as reference zero in eV (${E_{\rm f}}={E_{\rm {TMNW}}}-{E_{\rm {TMCp_2}}}+2{E_{\rm H}}$, in which $E_{\rm {TMNW}}$, $E_{\rm {TMCp_2}}$ and $E_{\rm H}$ denote the energies of metallocene nanowire, corresponding metallocene molecule and H atom, respectively), energy difference between ferromagnetic and antiferromagnetic configuration in meV ($\Delta E={E_{\rm {FM}}}-{E_{\rm {AFM}}}$), local magnetic moment in $\mu_{\rm B}$ ($M$) on the metal atom, one-electron gap by PBE in eV ($E_{\rm g}$), lowest exciton energy in eV ($\varepsilon_0$), corresponding binding energy in eV ($E_{\rm b}$), and excitonic instability or not (EI) for three metallocene nanowires. Spin-resolved results are given for CrNW, and $\uparrow$ and $\downarrow$ denote the results for spin-up and spin-down channels, respectively.}
	\label{tab:group}
	\begin{ruledtabular}
		\begingroup
		\renewcommand{\arraystretch}{1.5}
		\begin{tabular}{ccccccccccccc}
			
			& \emph{c}  & $E_{\rm f}$ & $\Delta E$  & $M$  & $E_{\rm g}$  & $\varepsilon_0$ &  $E_{\rm b}$ & EI\\			
			\hline
			
			ScNW & 5.58 & -0.05  & 55  & 1  & 0.46 & -0.71 & 2.24 & yes \\
			
			CrNW & 5.38 & 0.01  & -33 & 2 & 2.23$\uparrow$ 0.48$\downarrow$ & 1.71$\uparrow$ -0.37$\downarrow$ & 2.20$\uparrow$ 3.01$\downarrow$ & no$\uparrow$ yes$\downarrow$\\
			
			FeNW & 5.17 & 0  & --- & 0  & 2.17 & 2.29 & 1.88 & no\\
		\end{tabular}
		\endgroup
	\end{ruledtabular}
\end{table*}

\begin{figure*}[tbp]
	\includegraphics[width=1\columnwidth]{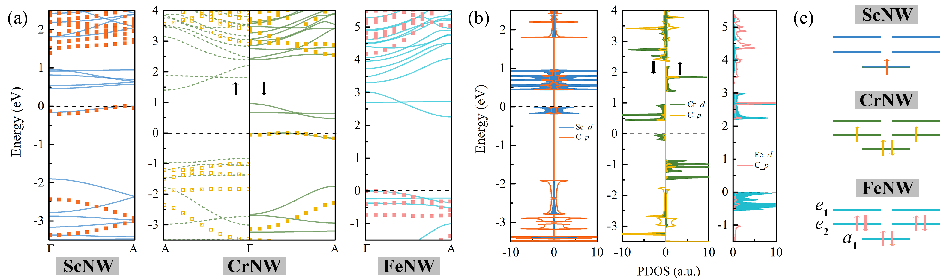}
	\caption{\label{Fig3} (Color online) (a) Band structures of ScNW, CrNW and FeNW by PBE (solid and dashed lines) and $G_0W_0$ (solid and empty squares), respectively. (b) Projected density of states (PDOS) on transition-metal $d$-orbitals and C $p$-orbitals for ScNW, CrNW and FeNW. In (a) and (b), the Fermi levels are set to zero (dashed lines). Bands and PDOS are spin-resolved for ScNW and CrNW, while not for FeNW. For ScNW, the bands for two spin channels are exactly degenerate. (c) Characterized transition-metal $d$-electron configurations for ScNW, CrNW and FeNW. $\uparrow$ and $\downarrow$ represent an occupied electron in the spin-up and spin-down channels, respectively.}
\end{figure*}

Table I summarizes the calculated results for the three TMNWs. Structurally, the lattice constant of TMNWs decreases as the metal atomic number increases from Sc to Cr to Fe. Energetically, the formation energy of CrNW is comparable to that of the experimentally synthesised FeNW, while that of ScNW is lower by 0.05 eV per metal atom. Thus, from a thermodynamic point of view, it is also feasible to fabricate ScNW and CrNW in a similar way to FeNW.

The electronic structures of ScNW, CrNW and FeNW are plotted in Figs. 3(a) and 3(b). All the three features indirect-gap semiconductors by PBE. Both ScNW and CrNW are spin-polarized, unlike the non-spin-polarized FeNW. Specifically, ScNW is antiferromagnetic with an energy 55 meV/Sc lower than that of the ferromagnetic configuration. Instead, CrNW is ferromagnetic with an energy 33 meV/Cr lower than that of the antiferromagnetic configuration.

As shown in Fig. 3(a), ScNW has a minimum PBE gap of 0.46 eV, with the valence band maximum and conduction band minimum located at $A$ and $\Gamma$ respectively. $G_0W_0$ increases the gap to 1.43 eV, accompanied by a slight shift of the valence band maximum from $A$ towards $\Gamma$. CrNW has PBE gaps of 2.23 eV and 0.48 eV for the spin-up and spin-down channels, respectively. They are remarkably increased to 3.91 eV and 2.50 eV by $G_0W_0$. FeNW shows a PBE gap of 2.17 eV and a $G_0W_0$ one of 4.17 eV.

From the projected density of states shown in Fig. 3(b), one can see that the states near the Fermi energy are mainly contributed by transition-metal $d$-orbitals. It is thus possible to understand the electronic and magnetic properties of different nanowires in terms of their characteristic $d$-electron configurations\cite{Shen,LiJPC}. According to the H\"{u}ckel rule, the Cp, as a five-electron metastable radical, tends to capture an extra electron to form a stable aromatic configuration\cite{Shen}. Since the unit-cell of TMNWs contains two Cp rings, this results in the neighbouring metal atom losing its two 4$s$ electrons. Under an approximately local environment of TMCp$_2$, five TM 3$d$ orbitals split into $a_1$ singlet, two-fold degenerate $e_2$ and $e_1$ doublets\cite{LiYC}. As a result, Sc($d^1$), Cr($d^4$), and Fe($d^6$) form the characteristic valence-electron populations shown in Fig. 3(c).

With this, it is immediately obvious that ScNW and CrNW are magnetic, with local magnetic moments on Sc and Cr atoms of 1 and 2 $\mu_{\rm B}$, respectively, while FeNW is not. These agrees very well with our first-principles calculations (see Table I) and are also in line with previous study\cite{DaiJMCC}. In addition, the population helps to understand the relative gap sizes of the different nanowires. Since the splitting of $a_1$ and $e_2$ is usually smaller than that of $e_2$ and $e_1$\cite{LiYC}, the gap of ScNW is smaller than that of FeNW. This is also evidenced in CrNW, where the calculated spin-up gap is much larger than the spin-down one for the same reason.

Owing to the selection rule, such $d$-$d$ gap nature determines a suppressed contribution to the system screening from the band-edge states\cite{Jianghf,BuXT}. Combined with their reduced dimension, TMNWs are expected to have very interesting excitonic effects, not only in terms of large binding energies, but also in terms of being decoupled from the one-electron gap, all of which are favourable for the occurrence of excitonic instabilities\cite{usEI,linear}. In this regard, we solve the BSE and obtain their low-energy exciton spectra as presented in Fig. 4. Note that only zero-momentum excitons are considered here. Despite the presence of an indirect gap, the near-Fermi-level bands are all relatively flat due to the $d$-orbital nature, so this treatment is expected to have little effect on the lower exciton energy levels.

\begin{figure}[tbp]
	\includegraphics[width=0.9\columnwidth]{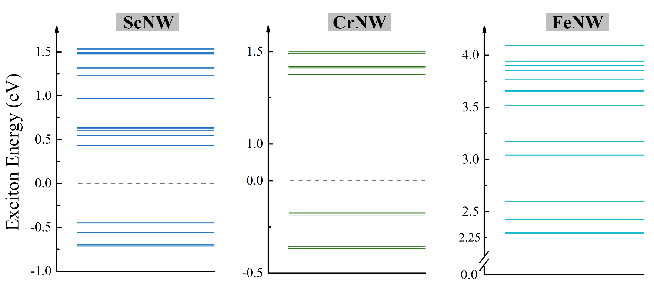}
	\caption{\label{Fig4} (Color online) Low-energy exciton levels by solving the BSE on top of $G_0W_0$ results for (a) ScNW, (b) CrNW and (c) FeNW. For CrNW, only the spectrum of spin-down channel is given and the lowest-energy level for spin-up is as high as 1.71 eV (see Table I).}
\end{figure}

The lowest exciton of ScNW has a negative energy of -0.71 eV, signalling the presence of excitonic instability. As usual, it is an optically inactive dark exciton\cite{usEI,usDong}. For an antiferromagnet, spin-triplet exciton has lower energy than spin-singlet one due to the exciton exchange interaction\cite{Jianghf,Mazzone}. Accordingly, the ScNW would form a spin-triplet excitonic insulator when spontaneously generated excitons condense. The two spins of CrNW have one-electron gaps of very different sizes, and their competition with exciton binding energies leads to essentially distinct results, yielding spin-resolved excitonic instabilities. As listed in Table I, the lowest exciton of the spin-up channel has a positive energy of 1.71 eV; on the contrary, it is negative, -0.37 eV, for the spin-down channel with a much smaller one-electron gap. Therefore, the CrNW is a half excitonic insulator with single-spin exciton condensate\cite{Jianghf}. From our first-principles $GW$-BSE calculations, the binding energies corresponding to spin-up and spin-down excitons are inferred to be 2.20 and 3.01 eV, respectively. Whilst, the FeNW is a usual semiconductor with a lowest exciton energy of 2.29 eV (see Table I and Fig. 4).

Both ScNW and CrNW are magnetic excitonic insulators, which provide an additional degree of freedom for probing spontaneous condensation of excitons. For example, spin-triplet exciton condensation can host spin superfluidity\cite{Jiangtri}. We also anticipate that the interplay between magnetism and exciton condensation would cause magnetic excitonic insulators to exhibit some unique features different from those of normal magnetic insulators, thus providing evidence for identification.

\begin{figure}[tbp]
	\includegraphics[width=1\columnwidth]{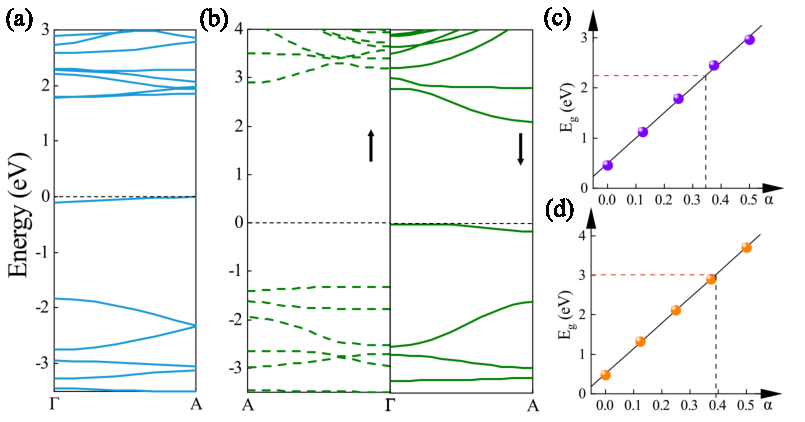}
	\caption{\label{Fig5} (Color online) Band structures of (a) ScNW and (b) CrNW by HSE06. The Fermi levels are set to zero (dashed lines). Gap size as a function of exact-exchange weights $\alpha$ for (c) ScNW and (d) CrNW. Note that only the spin-down values are shown for CrNW, since its large spin-up gap makes this channel independent of excitonic instability at all times. The red horizontal dashed line indicates the exciton binding energies in Table I, from which the critical $\alpha$ for excitonic instability is estimated.}
\end{figure}

In addition, both magnetic ground state and excitonic instability could in principle be the consequence of a semi-local approximation in the use of PBE exchange-correlation functional\cite{Barborini}. To clarify this pivotal point, we have further performed calculations using the HSE06 hybrid functional. We first perform structural relaxations on the supercell containing two transition-metal atoms. The results show that no Peierls distortion occurs. Then, we compare the energetics of different magnetic configurations. The results show that ScNW remains in the antiferromagnetic ground state with an energy 35 meV/Sc lower than that of the ferromagnetic configuration, while CrNW remains in the ferromagnetic ground state with an energy 10 meV/Cr lower than that of the antiferromagnetic configuration. Compared with the results of the PBE in Table I, HSE06 reduces the energy difference by 20 meV/Sc and 23 meV/Cr, respectively.

Finally, we calculate the one-electron bands at different exact-exchange weights, i.e., $\alpha$ = 0.125, 0.25, 0.375, and 0.5. We find that these bands resemble each other except for different gap sizes. In Figs. 5(a) and 5(b) we give typical bands for ScNW and CrNW at $\alpha$ = 0.25. They are similar to the ones by PBE and GW [see Fig. 3(a)]. Further, we present the gap size as a function of $\alpha$ in Figs. 5(c) and 5(d), respectively. Note that only the spin-down values are shown for CrNW, since its large spin-up gap makes this channel independent of excitonic instability at all times. As can be seen, the gap approximately exhibits a linear dependence on $\alpha$, namely, $E_{\rm g}$ = 5.06$\alpha + $ 0.49 and $E_{\rm g}$ = 6.42$\alpha + $0.50 for ScNW and CrNW, respectively. Since solving the BSE on top of HSE06 is very computationally expensive, we use the exciton binding energies in Table I [denoted by red horizontal dashed lines in Figs. 5(c) and 5(d)] as estimates. With this approach, we obtain that the excitonic instability of ScNW and CrNW remains robust if $\alpha <$ 0.35 and $\alpha <$ 0.39.

In summary, using statistical methods combined with first-principles calculations, we have investigated spontaneous Bose-Einstein condensation of excitons in the 1D systems. It is shown that 1D excitonic insulators do exist and their critical temperatures are even much higher than those of 2D and 3D excitonic insulators. However, 1D excitonic insulators have rarely been studied in the past, due in large part to the Hohenberg-Mermin-Wagner theorem ruling out its existence. We expect that the present study can revive interest in 1D excitonic insulators, thereby bringing new opportunities for the identification and future application of the excitonic insulator. Regarding the material realization, we predict that the single-chain staircase Scandocene and Chromocene wires are an antiferromagnetic spin-triplet excitonic insulator and a ferromagnetic half-excitonic insulator, respectively, providing a platform for exploiting the multifaceted nature of magnetism and spontaneous exciton condensate.

\begin{acknowledgments}
Y.L. thanks Zhirong Liu for fruitful discussions. This work was supported by the Ministry of Science and Technology of China (Grant Nos. 2023YFA1406400 and 2020YFA0308800) and the National Natural Science Foundation of China (Grant No. 12074034).
\end{acknowledgments}


\begin{thebibliography}{90}%
\makeatletter
\bibitem{Mott} N. F. Mott, The transition to the metallic state, Philos. Mag. \textbf{6}, 287 (1961).

\bibitem{Knox} R. S. Knox, in Solid State Physics, edited by F. Seitz and D. Turnbull (Academic Press, New York, 1963), Suppl. 5, p. 100.

\bibitem{Kohn} D. J\'{e}rome, T. M. Rice, and W. Kohn, Excitonic insulator, Phys. Rev. \textbf{158}, 462 (1967).

\bibitem{Rice} B. I. Halperin and T. M. Rice, Possible anomalies at a semimetal-semiconductor transistion, Rev. Mod. Phys. \textbf{40}, 755 (1968).

\bibitem{Eisenstein} J. P. Eisenstein and A. H. MacDonald, Bose-Einstein condensation of excitons in bilayer electron systems, Nature \textbf{432}, 691 (2004).

\bibitem{Jiangtri} Z. Y. Jiang, W. K. Lou, Y. Liu, Y. C. Li, H. F. Song, K. Chang, W. H. Duan, and S. B. Zhang, Spin-triplet excitonic insulator: The case of semihydrogenated graphene, Phys. Rev. Lett. \textbf{124}, 166401 (2020).

\bibitem{Sham} T. Portengen, T. \"{O}streich, and L. J. Sham, Linear and nonlinear optical characteristics of the Falicov-Kimball model, Phys. Rev. Lett. \textbf{76}, 3384 (1996).

\bibitem{Batista} C. D. Batista, Electronic ferroelectricity in the Falicov-Kimball model, Phys. Rev. Lett. \textbf{89}, 166403 (2002).

\bibitem{Jianghf} Z. Y. Jiang, Y. C. Li, W. H. Duan, and S. B. Zhang, Half-excitonic insulator: A single-spin Bose-Einstein condensate, Phys. Rev. Lett. \textbf{122}, 236402 (2019).

\bibitem{Liufeng} G. Sethi, Y. Zhou, L. Zhu, L. Yang, and F. Liu, Flat-band-enabled triplet excitonic insulator in a diatomic kagome lattice, Phys. Rev. Lett. \textbf{126}, 196403 (2021).

\bibitem{Mazzone} D. G. Mazzone, Y. Shen, H. Suwa, G. Fabbris, J. Yang, S.-S. Zhang, H. Miao, J. Sears, Ke Jia, Y. G. Shi, M. H. Upton, D. M. Casa, X. Liu, J. Liu, C. D. Batista, and M. P. M. Dean, Antiferromagnetic excitonic insulator state in Sr$_3$Ir$_2$O$_7$, Nat. Commun. \textbf{13}, 913 (2022).

\bibitem{Durr} L. Du, X. Li, W. Lou, G. Sullivan, K. Chang, J. Kono, and R. R. Du, Evidence for a topological excitonic insulator in InAs/GaSb bilayers. Nat. Commun. \textbf{8}, 1971 (2017).

\bibitem{Daniele} D. Varsano, M. Palummo, E. Molinari, and M. Rontani, A monolayer transition-metal dichalcogenide as a topological excitonic insulator, Nat. Nanotechnol. \textbf{15}, 367 (2020).

\bibitem{Wusanfeng} Y. Jia, P. Wang, C. Chiu, Z. Song, G. Yu, B. J\"{a}ck, S. Lei, S. Klemenz, F. A. Cevallos, M. Onyszczak, N. Fishchenko, X. Liu, G. Farahi, F. Xie, Y. Xu, K. Watanabe, T. Taniguchi, B. A. Bernevig, R. J. Cava, L. M. Schoop, A. Yazdani, and S. Wu, Evidence for a monolayer excitonic insulator, Nat. Phys. \textbf{18}, 87 (2022).

\bibitem{Cobden} B. Sun, W. Zhao, T. Palomaki, Z. Fei, E. Runburg, P. Malinowski, X. Huang, J. Cenker, Y. T. Cui, J. H. Chu, X. Xu, S. S. Ataei, D. Varsano, M. Palummo, E. Molinari, M. Rontani, and D. H. Cobden, Evidence for equilibrium exciton condensation in monolayer WTe$_2$, Nat. Phys. \textbf{18}, 94 (2022).

\bibitem{DongTEI} S. Dong and Y. C. Li, Robust high-temperature topological excitonic insulator of transition-metal carbides (MXenes), Phys. Rev. B \textbf{107}, 235147 (2023).

\bibitem{LiXZ} H. Yang, J. Zeng, Y. Shao, Y. Xu, X. Dai, and X. Z. Li, Spin-triplet topological excitonic insulators in two-dimensional materials, Phys. Rev. B \textbf{109}, 075167 (2024).

\bibitem{usEI} Z. Y. Jiang, Y. C. Li, S. B. Zhang, and W. H. Duan, Realizing an intrinsic excitonic insulator by decoupling exciton binding energy from the minimum band gap, Phys. Rev. B \textbf{98}, 081408(R) (2018).

\bibitem{usDong}S. Dong and Y. C. Li, Transition from band insulator to excitonic insulator via alloying Se into monolayer TiS$_3$: A computational study, Phys. Rev. B \textbf{102}, 155119 (2020).

\bibitem{DongAlSb} S. Dong and Y. C. Li, Excitonic instability and electronic properties of AlSb in the two-dimensional limit, Phys. Rev. B \textbf{104}, 085133 (2021).

\bibitem{LxzNJP} H. Yang, X. Wang, and X. Z. Li, A scenario for high-temperature excitonic insulators, New J. Phys. \textbf{24}, 083010 (2022).

\bibitem{Hohenberg} P. C. Hohenberg, Existence of long-range order in one and two dimensions, Phys. Rev. \textbf{158}, 383 (1967).

\bibitem{Mermin} N. D. Mermin and H. Wagner, Absence of ferromagnetism or antiferromagnetism in one-or two-dimensional isotropic Heisenberg models, Phys. Rev. Lett. \textbf{17}, 1133 (1966).

\bibitem{Walker} M. B. Walker, Nonexistence of excitonic insulators in one and two dimensions. Can. J. Phys. \textbf{46}, 817 (1968).

\bibitem{Gong} C. Gong, L. Li, Z. Li, H. Ji, A. Stern, Y. Xia, T. Cao, W. Bao, C. Wang, Y. Wang, Z. Q. Qiu, R. J. Cava, S. G. Louie, J. Xia, and X. Zhang, Discovery of intrinsic ferromagnetism in two-dimensional van der Waals crystals, Nature (London) \textbf{546}, 265 (2017).

\bibitem{Huang} B. Huang, G. Clark, E. Navarro-Moratalla, D. R. Klein, R. Cheng, K. L. Seyler, D. Zhong, E. Schmidgall, M. A. McGuire, D. H. Cobden, W. Yao, D. Xiao, P. Jarillo-Herrero, and X. Xu, Layer-dependent ferromagnetism in a van der Waals crystal down to the monolayer limit, Nature (London) \textbf{546}, 270 (2017).

\bibitem{RiceJPCM} M. J. Rice and Y. N. Gartstein, The excitonic ground state of the half-filled Peierls insulator, J. Phys.: Condens. Matter \textbf{17}, 4615 (2005).

\bibitem{Rice9} M. J. Rice, A class of polymeric semimetals that may exhibit quasi-one-dimensional excitonic insulator behavior, Synth. Met. \textbf{141}, 9 (2004).

\bibitem{Rice11} M. J. Rice and Y. N. Gartstein, Excitonic insulator transition in the conjugated polymer polyacene, Synth. Met. \textbf{141}, 11 (2004).

\bibitem{VarsanoNC} D. Varsano, S. Sorella, D. Sangalli, M. Barborini, S. Corni, E. Molinari, and M. Rontani, Carbon nanotubes as excitonic insulators, Nat. Commun. \textbf{8}, 1461 (2017).

\bibitem{Hellgren} M. Hellgren, J. Baima, and A. Acheche, Exciton Peierls mechanism and universal many-body gaps in carbon nanotubes, Phys. Rev. B \textbf{98}, 201103(R) (2018).

\bibitem{MnCp} J. Liu, G. -B. Liu, and Y. C. Li, Electric-field-driven excitonic instability in an organometallic manganese-cyclopentadienyl wire, Phys. Rev. B \textbf{104}, 085150 (2021).

\bibitem{CrBz} J. Liu and Y. C. Li, First-principles perspective on full-spectrum infrared photodetectors from doping an excitonic insulator, Phys. Rev. B \textbf{106}, 035135 (2022).

\bibitem{Barborini} M. Barborini, M. Calandra, F. Mauri, L. Wirtz, and P. Cudazzo, Excitonic-insulator instability and Peierls distortion in one-dimensional semimetals, Phys. Rev. B \textbf{105}, 075122 (2022).

\bibitem{Batista4} C. D. Batista, J. E. Gubernatis, J. Bon\v{c}a, and H. Q. Lin, Intermediate Coupling Theory of Electronic Ferroelectricity, Phys. Rev. Lett. \textbf{92}, 187601 (2004).

\bibitem{AFM} V. M. Santhini, O. Stetsovych, M. Ondr\'{a}\v{c}ek, J. I. Mendieta Moreno, P. Mutombo, B. de la Torre, M. \u{S}vec, J. Kl\'{\i}var, I. G. Star\'{a}, H. V\'{a}zquez, I. Star\'{y}, and P. Jel\'{\i}nek, On-Surface Synthesis of Polyferrocenylene and its Single-Chain Conformational and Electrical Transport Properties, Adv. Funct. Mater. \textbf{31}, 2006391 (2020).

\bibitem{PBE} J. P. Perdew, K. Burke, and M. Ernzerhof, Generalized Gradient Approximation Made Simple, Phys. Rev. Lett. \textbf{77}, 3865 (1996).

\bibitem{vasp} G. Kresse and J. Furthm\"{u}ller, Efficient iterative schemes for $ab initio$ total-energy calculations using a plane-wave basis set, Phys. Rev. B \textbf{54}, 11169 (1996).

\bibitem{PAW} P.E. Bl\"{o}chl, Projector augmented-wave method, Phys. Rev. B \textbf{50}, 17953 (1994).

\bibitem{PAW2} G. Kresse and D. Joubert, From ultrasoft pseudopotentials to the projector augmented-wave method, Phys. Rev. B \textbf{59}, 1758 (1999).

\bibitem{Hybertsen} M. S. Hybertsen and S. G. Louie, Electron correlation in semiconductors and insulators: Band gaps and quasiparticle energies, Phys. Rev. B \textbf{34}, 5390 (1986).

\bibitem{Rohlfing} M. Rohlfing and S. G. Louie, Electron-hole excitations and optical spectra from first principles, Phys. Rev. B \textbf{62}, 4927 (2000).

\bibitem{Shen} L. Shen, S.-W. Yang, M.-F. Ng, V. Ligatchev, L. P. Zhou, and Y. P. Feng, Charge-transfer-based mechanism for half-metallicity and ferromagnetism in one-dimensional organometallic sandwich molecular wires, J. Am. Chem. Soc. \textbf{130}, 13956 (2008).

\bibitem{LiJPC} Y. C. Li, G. Zhou, J. Li, J. Wu, B.-L. Gu, and W. H. Duan, Ab initio Study of Half-Metallicity and Magnetism of Complex Organometallic Molecular Wires, J. Phys. Chem. C \textbf{115}, 7292 (2011).

\bibitem{LiYC} Y. C. Li, X. B. Chen, G. Zhou, W. H. Duan, Y. Kim, M. Kim, and J. Ihm, Trends in charge transfer and spin alignment of metallocene on graphene, Phys. Rev. B \textbf{83}, 195443 (2011).

\bibitem{DaiJMCC} Y. Ma, Y. Dai, W, Wei, and B. Huang, Engineering intriguing electronic and magnetic properties in novel one-dimensional staircase-like metallocene wires, J. Mater. Chem. C \textbf{1}, 941 (2013).

\bibitem{BuXT} X. T. Bu and Y. C. Li, Optical signature for distinguishing between Mott-Hubbard, intermediate, and charge-transfer insulators, Phys. Rev. B \textbf{106}, L241101 (2022).

\bibitem{linear} Z. Y. Jiang, Z. R. Liu, Y. C. Li, and W. H. Duan, Scaling Universality between Band Gap and Exciton Binding Energy of Two-Dimensional Semiconductors, Phys. Rev. Lett. \textbf{118}, 266401 (2017).


\end{thebibliography}
\end{document}